\documentstyle[12pt,epsfig]{article}
 1
\def\as{\alpha_{\rm S}}

\def\gev{{\rm\, GeV}}
\def\citenum#1{{\def\@cite##1##2{##1}\cite{#1}}}
\def\citea#1{\@cite{#1}{}}

\def\as{\alpha_{\rm S}}

\def\gev{{\rm\, GeV}}
\def\gev2{{\,\mbox{GeV}^2}}

\def\b{\beta}
\def\a{\alpha}

\def\g{\gamma}

\def\l{\lambda}

\def\s{\sigma}

\def\t{\tau}

\def\({\left(}
\def\){\right)}

\def\citenum#1{{\def\@cite##1##2{##1}\cite{#1}}}
\def\citea#1{\@cite{#1}{}}

\def\l1vt{\vec{l_{1\perp}}}

\def\bt{b_{\perp}}

\def\bt2{$b^2_t$}

\def\jol1{$J_0(\,l_{1\perp}\,r_{\perp}\,)$}

\def\citea#1{\@cite{#1}{}}

\def\beq{\begin{equation}}
\def\eeq{\end{equation}}
\def\bea{\begin{eqnarray}}
\def\eea{\end{eqnarray}}

\def\eq#1{{Eq.~(\ref{#1})}}
\def\eqs#1#2{{Eqs.~(\ref{#1})--(\ref{#2})}}
\def\ds{\displaystyle}
\def\supinput{\mbox{\scriptsize input}}
\def\supDGLAP{\mbox{\scriptsize DGLAP}}
\def\df2dlnq2{\frac{\partial F_2(x,Q^2)}{\partial \ln Q^2}}
\def\dlnf2dln1x{\frac{\partial \ln  F_2(x,Q^2)}{\partial \ln (1/x)}}
\def\bbbz{{\mathchoice {\hbox{$\sf\textstyle Z\kern-0.4em Z$}}
{\hbox{$\sf\textstyle Z\kern-0.4em Z$}}
{\hbox{$\sf\scriptstyle Z\kern-0.3em Z$}}
{\hbox{$\sf\scriptscriptstyle Z\kern-0.2em Z$}}}}
\def\npb#1#2#3{    {\it Nucl. Phys. }{\bf B#1} (19#2) #3}
\def\plb#1#2#3{    {\it Phys. Lett. }{\bf B#1} (19#2) #3}
\def\prd#1#2#3{    {\it Phys. Rev. }{\bf D#1} (19#2) #3}

\def\prl#1#2#3{    {\it Phys. Rev. Lett. }{\bf #1} (19#2) #3}

\def\zpc#1#2#3{    {\it Z. Phys. }{\bf C#1} (19#2) #3}

\def\sjnp#1#2#3{   {\it Sov. J. Nucl. Phys. }{\bf #1} (19#2) #3}

\relax
\begin{document}
\newcounter{savefig}
\newcommand{\alphfig}{\addtocounter{figure}{1}%
\setcounter{savefig}{\value{figure}}%
\setcounter{figure}{0}%
\renewcommand{\thefigure}{\mbox{\arabic{savefig}-\alph{figure}}}}
\newcommand{\resetfig}{\setcounter{figure}{\value{savefig}}%
\renewcommand{\thefigure}{\arabic{figure}}}

\begin{titlepage}
\noindent
\begin{flushright}
\parbox[t]{8em}{
August 1998 \\
DESY-98-102\\ 
TAUP 2515/98\\
{\tt hep-ph/9808257}}
\end{flushright}
\vspace{1cm}
\begin{center}
{\Large \bf  THE  EFFECT OF SCREENING }\\[2ex]
{\Large \bf  ON THE $\mathbf x $ AND $\mathbf Q^2$ BEHAVIOUR}\\[2ex]
{\Large \bf  OF $\mathbf{F_2}$ SLOPES }
 \\[4ex]

{\large E. ~G O T S M A N${}^{a)\,1)}$,\,\,\,\,\,\,\, E. ~L E V I
N${}^{a)\,b)\,2)}$,}

{\large 
  U. ~M A O R${}^{a)\,3)}$\,\,\, and \,\,\,E. ~N A F T A L I$^{a)\, 4)}$}
\footnotetext{$^{1)}$ Email: gotsman@post.tau.ac.il .}
\footnotetext{$^{2)}$ Email: leving@post.tau.ac.il .}
\footnotetext{$^{3)}$ Email: maor@post.tau.ac.il .}
\footnotetext{$^{4)}$ Email: erann@post.tau.ac.il .}
\\[4.5ex]
{\it a)\,\,\, School of Physics and Astronomy}\\
{\it Raymond and Beverly Sackler Faculty of Exact Science}\\
{\it Tel Aviv University, Tel Aviv, 69978, ISRAEL}\\[1.5ex]
{\it b)\,\,\, DESY Theory Group}\\
{\it 22603, Hamburg, GERMANY}\\[2.5ex]
\end{center}
~\,\,
\vspace{1cm}

{\samepage
{\large \bf Abstract:}
A systematic study of $\frac{\partial F_2(x,Q^2)}{\partial \ln Q^2}$ and
$\frac{\partial \ln  F_2(x,Q^2)}{\partial \ln (1/x)}$ is carried out
in pQCD taking screening corrections into account. The result of
calculations, which are different from the non screened DGLAP
prediction, are compared  and shown to agree   
with the available experimental data as well as a pseudo data base
generated from the ALLM'97 
parameterization. This pseudo data base allows us  to study in detail our
predictions over a wider kinematic region than is available
experimentally, and allows us to make suggestions for future experiments.
Our results are compared with the GRV'94 parameterization (which is used as
an input for our calculations) as well as the recently proposed MRST
structure functions.

}
\end{titlepage}

\section{Introduction}
\par HERA  data  on $Q^2$ and $x$ dependences  of
  $\frac{\partial F_2}{\partial \ln Q^2}$, the logarithmic $Q^2$
derivative of the proton structure function $F_2(x,Q^2)$, have been
 published recently \cite{DATA1} \cite{DATA2}.
The behaviour of   $\frac{\partial F_2}{\partial \ln Q^2}$ is of
particular interest in the small $x$ limit of deep inelastic scattering (
DIS ), where the DGLAP evolution equations \cite{DGLAP} imply a 
relation
\beq \label{1}
  \frac{\partial F_2}{\partial \ln Q^2}\,\,\,=\,\,\frac{2 \as}{9
\,\pi}\,x\,G^{\supDGLAP}(x,Q^2)\,\,.
\eeq
\par
The most recent ZEUS data \cite{DATA2} are shown in Fig.1 together with
the corresponding GRV'94 \cite{GRV} predictions. Each of these data points
correspond 
to a different   $x$ and $Q^2$ value. The ( $ x, Q^2 $ ) of Fig.1
are averaged  values obtained from each of the experimental data
distribution 
bins. The wide spread of these ( $ x, Q^2 $ ) sets reflects the
constraints imposed by the availability of  very limited statistics in
the limit of very small $x$.

\begin{figure}[htbp]
\epsfig{file=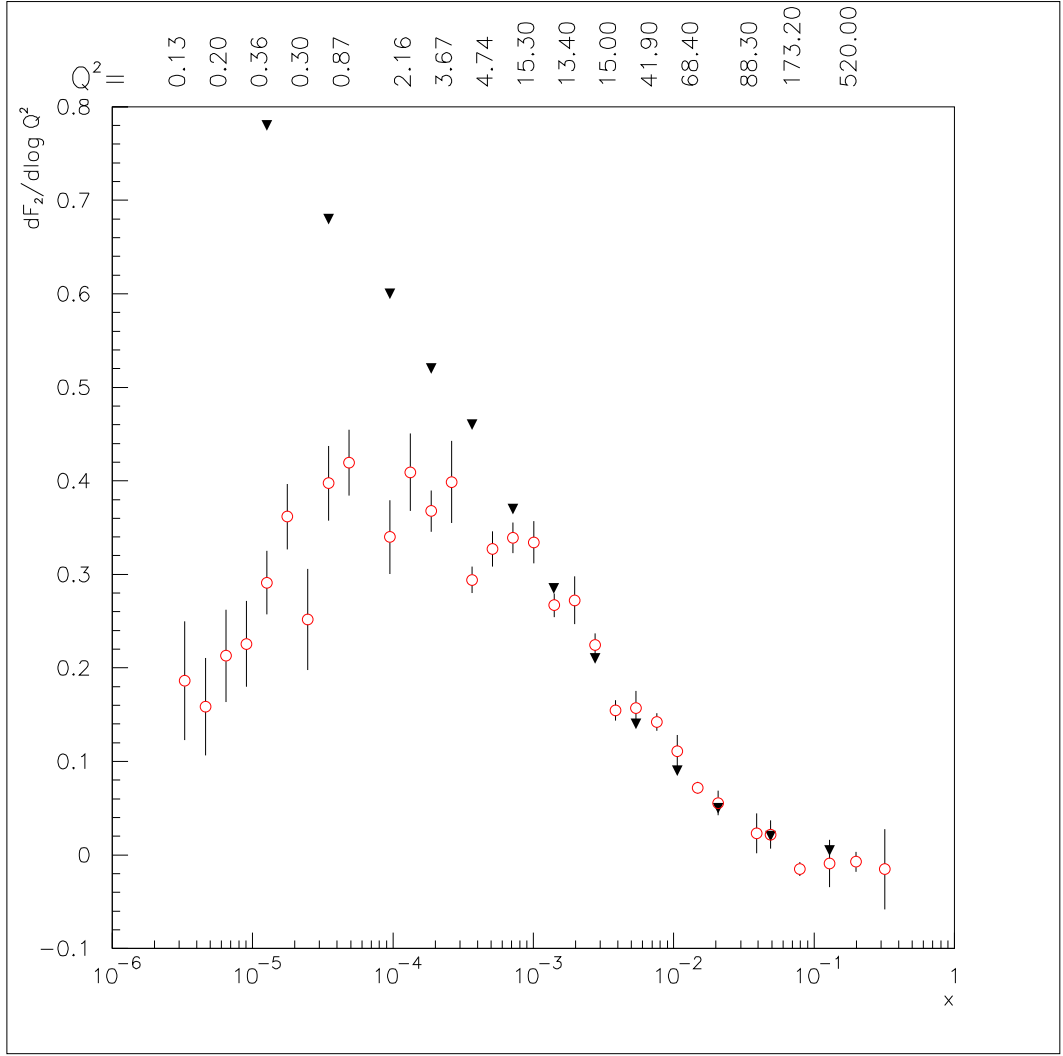,width=140mm}
\caption[The HERA data and GRV'94 prediction on the $F_2(x,Q^2)$
slope.]{\parbox[t]{0.80\textwidth}{The  HERA data and GRV'94
    predictions (triangles) on the $F_2(x,Q^2)$ slope. The data are
    taken from Ref.\cite{DATA2}.}}
\label{Fig.1} 
\end{figure}
\par As can be seen,   $\frac{\partial F_2}{\partial \ln Q^2}$  rises 
steeply with $\frac{1}{x}$ up to $x$ of approximately $ 5 \times 10^{-3}$,
for  $Q^2\,>\,5\,\gev2$. This is in agreement  with the prediction
of perturbative QCD \,\,( pQCD ) as manifested by the GRV'94
parameterization
of $F_2 (x, Q^2 )$. The data shows a dramatic departure  from this
prediction in the exceedingly small $x \,<\,5 \times 10^{-3}$ and low
$Q^2 \,<\,5\, \gev2 $. In this kinematic domain, while GRV'94 predicts a
continued increase of   $\frac{\partial F_2}{\partial \ln Q^2}$ with
$\frac{1}{x}$, the actual data indicates that  the logarithmic $Q^2$ slope
of $F_2$
 decreases significantly  with $\frac{1}{x}$,  up to the present
experimental limit of $x\,\simeq \,3 \times 10^{-6}$.

\par There are some obvious experimental and theoretical observations and  
questions prompted  by the above data:
\begin{enumerate}

\item  What is the detailed $x$ and $Q^2$ dependence of   $\frac{\partial
F_2}{\partial \ln Q^2}$ ? As observed, the phenomenon of decreasing
$\frac{\partial F_2}{\partial \ln Q^2}$ is associated with small $Q^2$ and
exceedingly small $x$. However, since the data is so sparse  it
is difficult to deduce   its detailed behaviour . In particular, the
available data does not provide us with  a clue as to the conduct  of 
  $\frac{\partial F_2}{\partial \ln Q^2}$   at  medium and high $Q^2$,
in the exceedingly small $x$ domain. Note, that the entire data sample 
with $x \,<\,5 \times 10^{-3}$ has $Q^2$ values smaller than $5\,\gev2$.

\item The experimental behaviour of  $\frac{\partial F_2}{\partial \ln
Q^2}$ as a function of $x$ and  $Q^2$  shown in Fig.1,
indicates a possible departure from the DGLAP pQCD behaviour of \eq{1},
for sufficiently small $x$.
An interesting  interpretation  in this direction ( MRST ) has been
recently 
proposed \cite{MRST}
 assuming  an initial  ``valence"  gluon input
distribution at small $Q^2$, which changes  rapidly  due to
QCD
evolution  to the conventional ``sea" distribution, at larger values of
$Q^2$.       We shall comment on this below.

The experimental  $F_2(x,Q^2)$      data  in the same kinematic region do
not show  any unconventional behaviour,
 i.e. $F_2$ is not sensitive enough, in
the very small $x$ domain, to resolve new features.
The new  data suggest   that the logarithmic $Q^2$ slope of $F_2$, 
is a more sensitive  measure  of  possible deviations 
from the conventional  pQCD dynamics. 

\item  The unexpected  phenomenon occurs in the transition region between
``hard" and ``soft" physics and one should carefully assess our ability to
understand it  within the framework of  pQCD. Specifically, the
predominant
``soft"
inclusive total real photoproduction cross section is experimentally
finite. Namely, the    $Q^2 = 0$  limit of $
\sigma_{tot}\,=\,\frac{4 \pi^2\,\alpha_{em}}{Q^2}\,F_2(x,Q^2)$ is finite
at any $W $ and thus $F_2 \,\,\propto\,\,Q^2$ as $ Q^2\,\rightarrow \,0$.
For a parameterization of the transition from DIS to real photoproduction,
see Ref.\ \cite{ALLM}. However, the scale of this transition is not
theoretically or experimentally specified.
Generally, it is  not clear if the observed change in the $x$ and $Q^2$
dependences
of  $\frac{\partial F_2}{\partial \ln Q^2}$ can be analyzed utilizing pQCD
techniques, or if this is just a signature for the dominance of ``soft"
non
perturbative physics, when both $Q^2$ and $x$ become small enough.
Regardless, of its detailed dynamics, the data suggest a relatively smooth
transition from the ``hard" to the ``soft" domains. Accordingly,  we 
 wish to match       the ``soft" ( non
perturbative
)
and the ``hard" ( perturbative ) contributions probably around
$Q^2\,\approx\,1  - 2 \,\gev2$, so as to better understand the ``soft"
limit of pQCD.

\item  The theoretical interpretation of the  new data presents a dual
challenge . On  one hand, it is desirable to understand  and
 formulate the dynamics responsible for the changed behaviour of
$\frac{\partial F_2}{\partial \ln Q^2}$, and  to comprehend  its
implications for pQCD. On the other hand, regardless of this dynamics, the
observed behaviour of   $\frac{\partial F_2}{\partial \ln Q^2}$ implies
that $xG(x,Q^2)$  differs from our previous   expectations in the limit of
very small $x$ and $Q^2$. This is bound to impose   changes on  the
input parton distributions which are currently used
\cite{GRV,MRS,CTEQ}. This is, indeed, the main point of Ref.\ \cite{MRST}.

\item An additional interesting study is to examine $\lambda \,=\,
\frac{\partial \ln F_2}{\partial \ln (1/x) }$ from which one  determines
the
dependence of $F_2$ as a power of $x$ at low  $x$ and  given $Q^2$,
$F_2\,\propto\,x^{
- \lambda(Q^2)}$.
$\lambda$ provides information pertinent to both Regge and pQCD analyses
of $F_2$. As such we shall be interested in its $Q^2$ and $x$  dependences,
and limits, and the constraints imposed on $\lambda$ by unitarity
\cite{GLR}.

\end{enumerate}
\par The goal of this paper is to expand and 
develop further the phenomenology and data analysis presented in our
recent
 letter \cite{GLMSLOPEF}   in which the modified  behaviour of
 $\frac{\partial 
F_2}{\partial \ln Q^2}$ was associated with the onset of unitarity 
screening   corrections ( SC ).For an early suggestion that
 $\frac{\partial F_2}{\partial\ln Q^2}$ is sensitive to SC , see Ref.
\cite{BCF}.

 To this end we study below the $x$ and $Q^2$ behaviour of both
$\frac{\partial F_2}{\partial\ln Q^2}$ and $\frac{\partial \ln
F_2}{\partial\ln(1/x)}$, checking the applicability of pQCD in $x$ and
$Q^2$ kinematic domains, not explored in detail previously.
         
\par A reliable execution of this program depends on our ability to
compare our calculations with the relevant data. As noted, such data
is available only at a few isolated values of averaged $x$ and
$Q^2$. This limited data base is not sufficient to test our suggested
phenomenology. To overcome this difficulty, we propose to use the
ALLM'97 computer generated data base which we call ``pseudo data ".
The ALLM parameterization \cite{ALLM} for the proton DIS structure
function is a 23 parameter description which provides an excellent
reproduction of the $\sigma_{tot} ( \gamma^* p ) $ data points. The
updated fit \cite{ALLM97} to 1356 data points with $ 0
\,\leq\,Q^2\,\leq\,5000\,\gev2$ and $ 3 \times 10^{-6}\,<\,x\,<\,0.85$
yields a $\chi^2/ndf$ of 0.97. Notably, ALLM'97 reproduces well both
the recent data on $\frac{\partial F_2}{\partial \ln Q^2}$ and
$\frac{\partial \ln F_2}{\partial \ln(1/ x)}$ (this is shown later in
Fig.\ref{Fig.5}b). 
As such, we assume that ALLM'97 provides an accurate and reliable
reproduction of the data, against which we can assess our calculations
and determine our free parameters.

\par The program of this paper is as follows: In  Section II we briefly
review the ALLM parameterization and its 1997 update. Section III is
devoted to a theoretical review of the SC 
in hard  parton physics and its relevance to $F_2$,
$\frac{\partial F_2}{\partial \ln Q^2}$ and  $\frac{\partial \ln
F_2}{\partial \ln (1/ x )}$ calculations. Our review examines
the SC in the quark and gluon sectors. The systematics of
$\frac{\partial F_2}{\partial \ln Q^2}$ and  $\frac{\partial \ln 
F_2}{\partial
\ln ( 1/x ) }$ and its comparison to the pseudo
data, defined above, are discussed in Section IV. Our summary and
conclusions  are given in Section V.

\section{ The  ALLM'97 Pseudo Data Base:}

The ALLM parameterization \cite{ALLM} was introduced in 1991 so as to
 reproduce $\s_{tot} ( \g^* p )$, the total
  $\g^* p $ cross section above the resonance region,  for a very wide
$Q^2$ range including real photoproduction ( $Q^2 = 0 $ ). This is a
multiparameter fit  ( 23 parameters  ) to all available data on
$\s_{tot} (\g p )$  and $F_2$, based on a  Regge - type approach
formulated so as 
to be compatible with pQCD and the DGLAP evolution  equations. In its
latest 1997 update  \cite{ALLM97}  a fit was performed to  1356
data points
\cite{DATA1,DATA3}, for which an excellent $\chi^2/ndf$ = 0.97 was obtained.

\par The ability to reproduce $\s_{tot} (\g^* p)$ over a wide
kinematic range of $ 3 \times 10^{-6}\,<\,x\,<\,0.85 $ and 
$0\,\leq\,Q^2\,\leq\,5000 \,\gev2 $, is further demonstrated by the
success of ALLM'97 in fitting the $F_2$ logarithmic slopes,
$\frac{\partial F_2}{\partial \ln Q^2}$ and 
$\frac{\partial \ln F_2}{\partial \ln (1/x )}$ well.  This is of
particular interest, as the parameterization reproduces the $F_2$ and
its logarithmic derivative data, also in the limit of exceedingly small
$x\, < \,5 \times 10^{-3}$.

\par The ALLM parameterization is presented by the following equations:
\beq \label{2}
F_2( x, Q^2 )\,\,\,=\,\,\,\frac{Q^2}{Q^2\,+\,m^2_0}\,\left(\,F^P_2( x,Q^2
)\,\,+\,\,F^R_2( x,Q^2 )\,\right)\,\,,
\eeq
where $F^P_2$ and $F^R_2$ are the respective contributions of the Pomeron
and the secondary Reggeon exchanges to $F_2$.
\begin{eqnarray}
&
F^P_2( x,Q^2 )\,\,\,=\,\,\,C_P(\t) (
\,\frac{1}{x_P}\,)^{\alpha_P(\t)\,-\,1}\,(\,1\,-\,x\,)^{\b_P(\t)}\,\,; &
\label{3}\\
&
F^R_2( x,Q^2 )\,\,\,=\,\,\,C_R(\t) (
\,\frac{1}{x_R}\,)^{\alpha_R(\t)\,-\,1}\,(\,1\,-\,x\,)^{\b_R(\t)}\,\,; &
\label{4}
\end{eqnarray}
where $\t\,\,=\,\,\ln \left(\frac{\ln \frac{Q^2  + Q^2_0}{\Lambda^2}}{\ln
\frac{ Q^2_0}{\Lambda^2}} \right)  $ is a slowly varying variable.

The two modified Bjorken $x$ - variables  are defined
\begin{eqnarray}
&
\frac{1}{x_P}\,\,\,=\,\,\,1\,\,+\,\,\frac{W^2 - M^2}{Q^2\,+\,m^2_P}\,\,;
& \label{5}\\
& \frac{1}{x_R}\,\,\,=\,\,\,1\,\,+\,\,\frac{W^2 - M^2}{Q^2\,+\,m^2_R}\,\,.
& \label{6}
\end{eqnarray}
$M$ denotes  the proton mass and $m_P$ and $m_R$ are fitted effective
Pomeron
and Reggeon scales. The scale parameters $m^2_0$,
$m^2_P$, $m^2_R$ and $Q^2_0$ control the smooth transition from $Q^2 = 0$
to high $Q^2$. When $Q^2\,\gg\,m^2_P$ and $  m^2_ R$, $x_P$ and $x_R$
approach
$x$. $C_R$, $\alpha_R$, $\b_R$ and $\b_P$ increase with $Q^2$ as
\beq \label{7}
f(\t)\,\,=\,\,f_1\,\,+\,\,f_2\,\t^{f_3}\,\,
\eeq
whereas $C_P$ and $\a_P$ decrease with $Q^2$ as
  \beq \label{8}
g(\t)\,\,=\,\,g_1\,\,+\,\, (
\,g_1\,\,-\,\,g_2\,)\,[\,\frac{1}{1\,+\,\t^{g_3}}\,\,-\,\,1\,]\,\,.
\eeq
\par  The parameterization depends on 23 parameters
which are determined from the data fit. About half of these parameters are
required to  describe the low $W$ ( high $x$ ) region where higher
QCD twist terms  are important. A specification  of the ALLM'97
fitted
parameters is given  in Table 2 of Ref. \cite{ALLM97}.

 In general, Eqs.~(\ref{3}) and (\ref{4}) were constructed to be
compatible with pQCD and reproduce two limiting cases:

(i) Reggeon - like behaviour at $x\,\,\rightarrow\,0$, with all
corrections to the simple one Reggeon exchange absorbed in the
dependence of the Reggeon intercept on $\t$\,.

(ii) The pQCD quark counting rules behaviour at $ x \,\rightarrow\,1
$, for which the power dependence of $\b_P$ and $\b_R$ on $\t$ are
expected.
 
In as much as the ALLM parameterization is constructed in a Regge-like
scheme, it is important to note, that it is not a theory to be
compared with the data, but a parameterization aimed at the best
possible reproduction of the data. The ALLM'97 parameters have two
basic features dictated by the data, but not expected in a simple
Regge-like theory.

~
 1) The Pomeron has a high scale of $m^2_P\,\simeq\,50\,\gev2$. This
high
value is necessary  to maintain a very smooth transition from
predominantly  ``soft" $Q^2\,=\,0$ to a  predominantly ``hard"  $Q^2$ of a
few $\gev2$. A smaller value of $m^2_P\,\simeq\,10\,\gev2$, suggested in
the original ALLM fit \cite{ALLM}, implies a very fast transition, which
is not compatible with the new  data.

2)  Theoretically, for a single Reggeon exchange,
$\alpha_R$ should  be independent of $Q^2$. The $Q^2$ dependence,
implied by the ALLM parameterization, secures the reproduction of the
DIS data, indicating the need to correct the simple Regge picture in DIS.

\par In this paper we evaluate the logarithmic derivatives
$\frac{\partial F_2}{\partial \ln Q^2}$ and 
$\frac{\partial \ln F_2}{\partial \ln (1/ x)}$ 
derived from ALLM , \eq{2} - \eq{8}. The
calculated $\frac{\partial F_2}{\partial \ln Q^2}$ and $\frac{\partial
\ln F_2}{\partial \ln (1/x)}$ coincide with the experimental data in
the kinematic region where they were measured.  The ability to produce
this detailed pseudo data base enables us to achieve two goals:
\begin{enumerate}
\item To achieve a realistic reproduction of the behaviour of
$\frac{\partial F_2}{\partial \ln Q^2}$ and $\frac{\partial \ln
F_2}{\partial \ln (1/x)}$, over the complete $Q^2$ range, with a
special emphasis on the small $Q^2$ and exceedingly small $x$ domain.

\item In our search for a dynamical explanation for possible
deviations from the DGLAP pQCD expectations, we offer in section IV a
pQCD calculation modified by SC which reproduces both the real data
and the pseudo data points for $Q^2\,>\,0.4\,\gev2$, as well as .
\end{enumerate}

 A presentation and discussion of these issues will be given in
section IV.

\par An important item  concerning ALLM is the uncertainty of the fit.
An error calculation of a 23 parameter fit,where  many of the parameters
are 
highly
correlated, is non trivial. Based on the estimates of Ref.\cite{ALLM},
we assess the pseudo data errors to be  approximately 8 - 10\% . 
Over the kinematic region of interest the ALLM calculated errors are
smaller than the comparable experimental errors. In section IV, which
is devoted to the data analysis including the ALLM'97 pseudo data, we
quote both errors whenever possible.

\section{ The Calculation of $\mathbf  F_2 $ slopes with SC}

~

\par As stated in   Ref.\cite{GLMSLOPEF}, we claim that a calculation of
 $\frac{\partial  F_2}{\partial \ln Q^2}$ including SC effects accounts
well for the data shown in Fig.1.

In the SC calculation we distinguish between two contributions leading to
observed deviations from the DGLAP predictions of \eq{1}. These are:

\begin{enumerate}

\item SC in the quark sector, corresponding to the passage of a $ q \bar q
$ pair through the target\,.

\item The analogous SC in the gluon sector, as $xG( x, Q^2 )$  appears 
 in the  expression for the slope of
$F_2$, \eq{1}.
\end{enumerate}
\subsection{Screening in the quark sector}
  SC in the quark sector, calculated in the Eikonal
approximation,  were derived some time ago \cite{LERY87} \cite{MU90}
leading to an  extensive phenomenology \cite{REF11}. In our context we
have
\jot 0.6em
\begin{eqnarray}
\ds{
\frac{\partial F^{SC}_2(x,Q^2)}{\partial \ln (Q^2/Q^2_0)} 
}& = &
\ds{ 
\frac{Q^2}{3 \,\pi^2}\,
\int d b^2_t\,\,\{\,\,1\,\,\,-\,\,\,e^{- \kappa_Q(x,Q^2; b^2_t)}\,\,\}\,\,;}
\label{9}\\
\ds{
\kappa_Q(x,Q^2; b^2_t) 
}& = &
\ds{
\frac{2 \pi \as }{3\, Q^2}\,\Gamma(b_t)\,\,xG^{\supDGLAP}(x,Q^2 )} \label{10}
\end{eqnarray} \jot 0em
$Q^2_0$ is the photon virtuality scale from which we start the DGLAP
evolution. $\Gamma( b_t )$ is the two gluon non-perturbative form factor of
the target in impact parameter  representation.
\beq \label{11}
\Gamma(b_t)\,\,=\,\,\frac{1}{\pi}\,\,\int\,e^{- i (\vec{b}_t
\cdot\vec{q}_t)}\,F(t)\,d^2 q_t
\eeq
with $t = - q^2_t$ and $F(t)$ denoting the form factor in $t$ -
representation.

 The fact that the $b_t$- dependence  factorizes from the  $x$ and
$Q^2$  dependences was shown  in Ref. \cite{GLMPH} from the DGLAP
evolution equations, using the factorization theorem \cite{FT}. It should
be stressed that the  factorization  we use in \eq{10}  is only  valid
 for $t \,\leq\,Q^2_0 $. Therefore, we assume below that only such
 small values of $t$  are important in our integrals. This is the
 justification for 
simplifying  the $b_t$-dependence of \eq{11}, and using a Gaussian
parameterization for $\Gamma(b_t)$:
\beq \label{12}
\Gamma(b_t)\,\,=\,\,\frac{1}{R^2}\,e^{- \frac{b^2_t}{R^2}}\,\,.
\eeq
where $R^2$ is a fitted parameter. 

\begin{figure}[htbp]
\epsfig{file=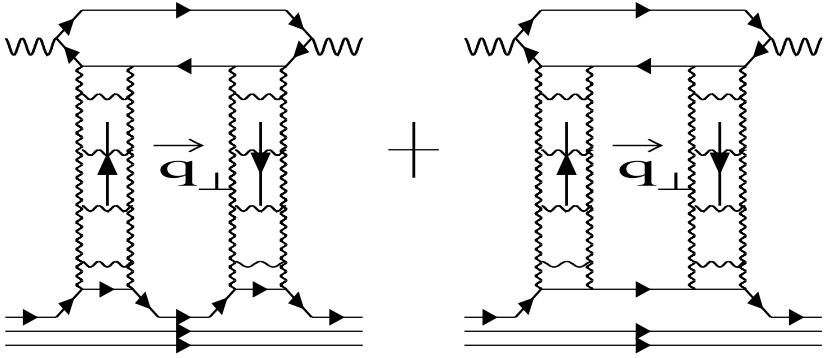,width=140mm,height= 65mm}
\caption{The first order SC $\propto \,\kappa^2$  for  $F_2(x,Q^2)$.}
\label{Fig.2}
\end{figure}

In Fig.2 we present the  lowest order SC to $F_2$ which  are  
proportional
to $\kappa^2_Q$,
   within the framework of the additive quark model ( AQM )
 ( see \eq{9} ).
 In the AQM we distinguish between two typical
scales in the
integration over $q_t$. In the first diagram of Fig.2 we integrate
over the distance between two constituent nucleon quarks. In the second
diagram of Fig.2 the relevant integration is over the size of the
constituent quark ( see Refs. \cite{AGLFR} \cite{GLM403} ). \eq{9}
assumes that the only constraint on the two gluons is their confinement
within a nucleon whose size is $R$.

The important property of \eq{9} is that the calculated SC in the quark
sector depends on a   distance $r^2_{\perp}\,\,=\,\,\frac{4}{Q^2}$ which
we consider to be short enough  for all values  of $Q^2$ in our
calculation.  Consequently, we assess the perturbative calculation of the
SC for 
$\frac{\partial F_2(x,Q^2)}{\partial \ln (Q^2/ Q^2_0)}$ to be  reliable.
This
is not so     for the calculation of the SC for $F_2$, 
which has an important
contribution from large distances $r^2_{\perp}\,>\,\frac{4}{Q^2}$ as will 
be discussed below.
As a result, the calculations of SC for $F_2$  are  not that well defined,
and
may
contain arbitrary  errors.

Following our previous publications \cite{GLMSLOPEF,GLMD}, we define a damping
factor
\beq \label{13}
D^2_Q( \kappa )\,\,=\,\,\frac{ \frac{\partial F^{SC}_2( x, Q^2)}{\partial
\ln(Q^2/Q^2_0)}}{ \frac{\partial F^{\supDGLAP}_2(x,Q^2)}{\partial\ln
Q^2}}\,\,,
\eeq
where $\frac{\partial F^{SC}_2( x, Q^2)}{\partial
\ln(Q^2/Q^2_0)}$ is calculated from  \eq{9}. The behaviour of
$D^2_Q(\kappa)$ as a function of  $x$ and $Q^2$  is plotted in
Fig.3a\, ,
where
we take
$R^2\,=\,10\,GeV^{-2}$.

\subsection{Screening in the gluon sector}
 The calculation of the SC in the gluon
sector using an  Eikonal  approach was derived by Mueller \cite{MU90}
and further discussed in
Ref.\cite{GLUONSC}. We obtain
\beq \label{14}
xG^{SC}(x,Q^2)\,\,=\,\,\frac{2}{
\pi^2}\,\int^1_x \,\frac{d x'}{x'} \,\int^{Q^2}_{Q^2_0} \,d Q'^2
\,\int
d b^2_t\,\,\{\,\,1\,\,\,-\,\,\,e^{- \kappa_G(x',Q'^2;
b^2_t)}\,\,\}\,\,+\,\,xG(x,Q^2_0)\,\,.
\eeq
Note that
$\kappa_G(x',Q'^2,b^2_t)\,\,=\,\,\frac{9}{4}\kappa_Q(x',Q'^2,b^2_t)$
defined in \eq{10}.

We can now define  a gluon damping factor 
\beq \label{15}
D^2_G(x,Q^2)\,=\,\frac{xG^{SC}(x,Q^2)}{xG^{\supDGLAP}(x,Q^2)}\,\,,
\eeq
where $x G^{SC} (x, Q^2) $ is calculated from \eq{14}.

A difficulty in our calculation stems from the fact that the $Q'^2$
integration spans not only the short ( perturbative ), but also the long (
non-perturbative ) distances. Lacking  solid theoretical estimates, we
suppress the long distance contributions  to \eq{14} by imposing a cutoff
on the  $Q'^2$ integration, so as to neglect the contributions    from
$r_{\perp}\,>\,\frac{1}{Q_0}$  integration. This { \it ad hoc } procedure
makes our estimate of $D^2_G (x,Q^2)$ less reliable than the   $D^2_Q
(x,Q^2)$
estimate. We have checked our cutoff procedure at  $Q^2_0\,=\,1\,\gev2$,
  using  the GRV'94
parameterization
for the solution of the DGLAP evolution equations and putting $xG(x,Q^2) =
0 $ for $Q^2 < Q^2_0 $ 
( we shall denote this model GLMN0 ) or imposing the GRV initial condition for $xG(x,Q^2)$ at $Q^2 = 
Q^2_0$ ( GLMN1 ). The output
result for
$D^2_G (x,Q^2)$   differs by about 10\%,  which lends credibility to our
calculation. A graphical representation of $D^2_G$ is provided in Fig.3b\, .
 Note that $D^2_G$ is consistently smaller than $D^2_Q$, i.e. SC in the
gluon sector are bigger than the SC in the quark sector. We also
observe that for any $Q^2$ the overall damping $D_Q^2\!\cdot\! D_G^2$
becomes significant once $x$ is small enough.

\begin{figure}[htbp]
\epsfig{file=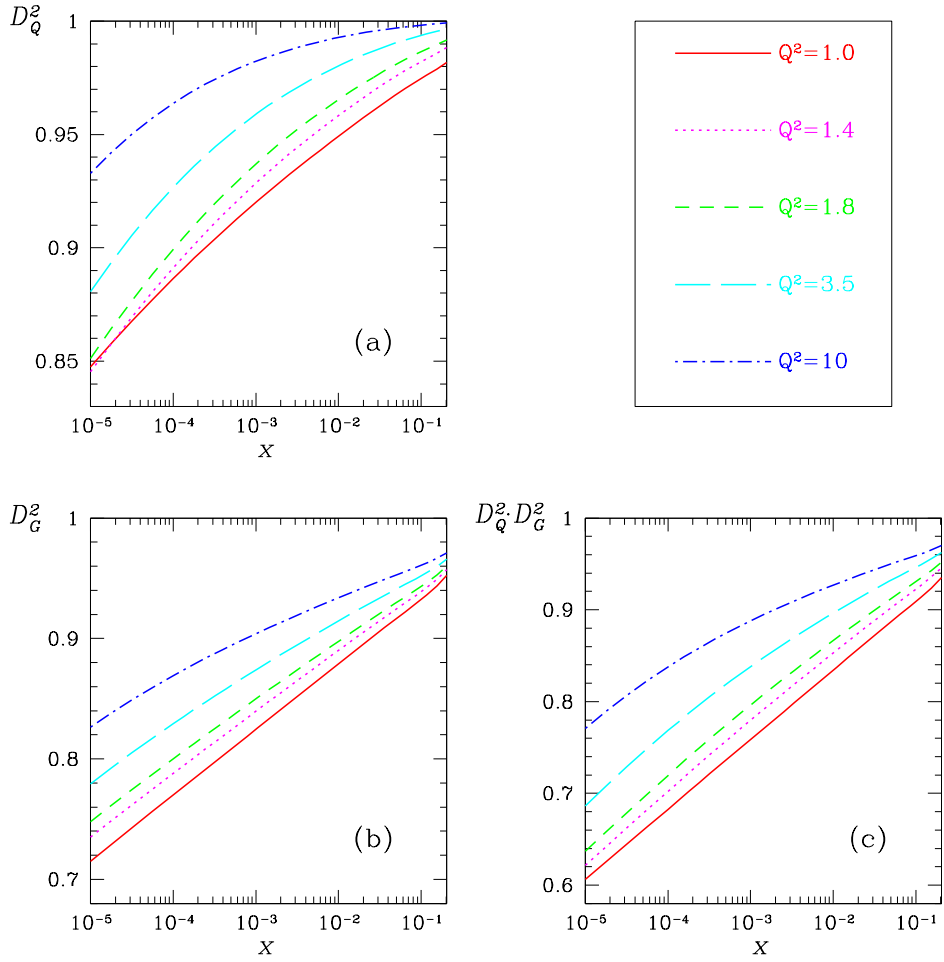,width=140mm}
\caption[The damping factors]{\parbox[t]{0.80\textwidth}{The damping
    factors: (a)~$D^2_Q(x,Q^2)$, 
             (b)~$D^2_G(x,Q^2)$ and \newline
             (c)~$D^2_Q(x,Q^2)\!\cdot\! D^2_G(x,Q^2)$.}}
\label{Fig.3}
\end{figure}

 To check our sensitivity to the region of low $Q'^2$ we use  a
general property of the gluon structure function, namely,  that
 gauge invariance requires    that $xG\,\propto\,Q^2$ at low $Q^2$\,\,\, 
(~see Ref.\cite{GLMSMD} for discussion ).\,\,\, Relying on this
general property we assume that
\begin{equation} \label{xglin}
 xG\left( x,Q^2 < \mu^2\right) = \frac{Q^2}{\mu^2} xG\left( x, \mu^2\right)\, ,
\end{equation}
where $\mu^2$ is a free parameter  in the range
$0.4\,\,\leq\,\,\mu^2\,\,\leq\,\,1\,\,\gev2$. We are  motivated by the
observation   that even    though GRV'94  evolves from a very low
$Q^2_0\,=\,0.4\,\gev2$, they actually fit the $F_2$ data  only   above
$Q^2$ of about $1 \,\gev2$. We have satisfied ourselves  that our
calculations are stable  ( within 10$\%$ ) to the choice of $\mu^2$.

\subsection{Overall screening}
Our final expression for $\frac{\partial F_2(x,Q^2)}{\partial \ln (Q^2/
Q^2_0)}$, i.e.  the $Q^2$ logarithmic slope of $F_2$ with SC included in
the calculation, is
\beq \label{16}
\frac{\partial F^{SC}_2(x,Q^2)}{\partial\ln(Q^2/Q^2_0)}
\,\,=\,\,D^2_Q(x,Q^2)\,D^2_G(x,Q^2)\,\frac{\partial
F^{\supDGLAP}_2(x,Q^2)}{\partial\ln(Q^2/Q^2_0)}\,\,.
\eeq
A detailed analysis of the systematics emerging from the above formula is
given in the next section.

 The choice of input parton distributions  obviously influences 
the  output. At first sight, we have a very clear prescription of what to do.
Indeed, the DGLAP evolution equations as well as our \eq{9} and \eq{14} (
see also the full set of equations below, \eq{17} - \eq{21}, where the
initial parton  distributions are noted  explicitly  ) are written
in  such a way that we depend on the initial parton distributions at a 
fixed
$Q^2
= Q^2_0$,  to solve them. These initial distributions, in principle,
include
everything that we have  from ``soft", non-perturbative physics.
All the information   relevant to  short distances
$r_{\perp}\,<\,\frac{1}{Q_0}$
is  included in
the above equations both with or without SC. 

In an ideal situation, our procedure should be to use as input a set of
parton distributions which have been determined at $Q^2 = Q^2_0$,
 without
the influence of data at $Q^2\,>\,Q^2_0$,  evolve  to higher $Q^2$ using
DGLAP evolution, and then correct for SC at   $Q^2\,>\,Q^2_0$,  utilizing
the damping factors.

Unfortunately,
such a set of input parton structure functions is not available, and we
must make do with one of the available parameterization.
 These sets are obtained  from global fits combined with DGLAP evolution
equations, i.e. the determination of the input is also influenced by
the
data with $Q^2\,>\,Q^2_0$ which may contain SC.  We have chosen
to work with  GRV'94  parameterization  since this is a relatively old
parameterization and
as such most of the fitted data has $x > 10^{-3}$. Our estimates show that
in this kinematic region we expect the SC to be small.

The GRV'94 parameterization has an additional advantage for us.  It
should be stressed that \eq{9}, \eq{14} and \eqs{17}{21} in the
following are proven in the double log approximation ( DLA ) in which
we consider both $\as\,\ln(1/x)\,>\,1$ and
$\as\,\ln(Q^2/Q^2_0)\,>\,1$. In the GRV'94 parameterization the
initial value of $Q^2_0\,=\,0.4\gev2$ is so small, that the DLA is
effectively valid above $Q^2\,\,\approx\,\,1\gev2$. Indeed, above
$1\gev2$ GRV'94 provides a good reproduction of the DIS data.

In a previous publication \cite{GLM403} we have discussed the
possibility that the proton is better described as a two radii object.
This description is based on the large experimental difference between
the slopes in $\frac{d \s}{d t}$ of $\g p \,\rightarrow\,J/\Psi p $
and $\g p \,\rightarrow\,J/\Psi X $. We have formulated this option in
Ref.  \cite{GLM403} and have applied it to $\frac{\partial
F^{SC}_2(x,Q^2)}{\partial\ln(Q^2/Q^2_0)}$ in Ref. \cite{GLMSLOPEF} in
which both a single and two radii models of the proton were
examined. In as much as we endorse the two radii idea, we are unable,
at this stage, to offer a decisive experimental signature that will
discriminate between the one and two radii models ,when dealing with
$F_2$ and its logarithmic slopes.  This being the case, we limit
ourselves in this paper to the simpler one radius approach.  A
detailed formalism for the two radii model was given in
Refs. \cite{GLM403} and \cite{GLMSLOPEF}.

\subsection{A screened calculation for $\mathbf \lambda ( Q^2 )$}
 We list below the expressions for $ F^{SC}_2( x, Q^2 )$ and  $ \lambda (
Q^2 )\,=\,
\frac{\partial \ln  F^{SC}_2(x,Q^2)}{\partial\ln(1/x)}$ which include
screening corrections.
The set of formulae are :
\jot 0.65em \begin{eqnarray}
F^{SC}_{2}( x, Q^2 ) &=& 
\frac{R^2\,N_c}{2\,\pi^2}\,\sum^{N_f}_{f=1}\,Z^2_f\,\,
\int^{Q^2}_{Q^2_0}\,d Q'^2\,
   \left\{\,C\,+\,
      \ln\kappa^{SC}_Q(x,Q'^2)\,+\,
      E_1\left(\kappa^{SC}_Q(x,Q'^2)\right)\,
   \right\} \nonumber\\
& & \mbox{} + F^{\supinput}_2(x,Q^2_0)\,\,; 
\label{17}\\
\frac{\partial F^{SC}_{2}( x, Q^2 ) }{ \partial \ln (1/x)} &=& 
\frac{R^2\,N_c}{2\,\pi^2}\,\sum^{N_f}_{f=1}\,Z^2_f\,\,
\int^{Q^2}_{Q^2_0}\,d Q'^2\,
\frac{\partial\ln \left(x G^{SC}\left(x,Q'^2\right)\right)}{\partial\ln (1/x)}
\,\left\{\,1\,\,-\,\,e^{ -\kappa^{SC}_Q(x,Q'^2)}\,
\right\} \nonumber\\
& & \mbox{} + \frac{\partial F^{\supinput}_2(x,Q^2_0)}{\partial\ln(1/x)}
    \,\,;
\label{18}\\
\kappa^{SC}_Q(x,Q^2) &=& \frac{2 \,\pi\,\as \left(Q^2\right)}{3\,R^2\,Q^2}\,x
G^{SC}(x,Q^2)\,\,; 
\label{19}\\
xG^{SC}\left(x, Q^2\right) &=& 
\frac{2 \,R^2}{\pi^2}\,
\int^1_x \,\frac{d x'}{x'}\,\int^{Q^2}_{Q^2_0}\,d \,Q'^2\,
      \left\{\,C\,+\,
      \ln \kappa^{\supDGLAP}_G(x',Q'^2)\,+\,
      E_1 \left( \kappa^{\supDGLAP}_G (x',Q'^2)\right)\,
   \right\} \nonumber\\ 
& & \mbox{} +  xG^{\supinput}(x,Q^2_0)\,\,;
\label{20}\\
\kappa^{\supDGLAP}_G(x,Q^2) &=& \frac{3\,\pi\,\as 
\left(Q^2\right)}{2\,R^2\,Q^2}\,xG^{\supDGLAP}(x,Q^2)\,\,.
\label{21}
\end{eqnarray} \jot 0em

We have to integrate over the all distances including the long
distance region. As we have discussed, this gives rise to undefined
errors in our calculation which we attempt to estimate, assuming
different cutoff values of the distances
$r_{\perp}\,\,>\,\,\frac{1}{Q_0}$.  All integrals over the impact
parameter $b_t$, were done using the Gaussian parameterization for the
$b_t$ dependence of \eq{12}. In the above set of equations $C$ denotes
the Euler constant (~$ C \,=\,0.577$~) and $E_1(x)\,=\,- Ei(-x)$ is
the exponential integral function (see Ref. \cite{MATH}).  $N_c$ and
$N_f$ are the colour and flavour degrees of freedom in QCD.  $Z_f$ is
the fraction of the electrical charge carried by a quark with flavor
$f$.

\subsection{Asymptotic predictions}
   It is instructive to discuss here
the asymptotic predictions of \eq{9} \,-\,\eq{21} in the region of very
small $x$. Such predictions have  been discussed previously  (~see Refs.
\cite{GLMSLOPEF,MU90,AGLFR,MUDIS98}~)  and we consider
them here  as  a limit with which we can compare our actual
calculations. 

In the region of small $x$,  which is defined in our formulae as a region
where  
\beq \label{22}
\kappa^{\supDGLAP}_Q(x,Q^2)\,\,=\,\,\frac{2 \,\pi\,\as (Q^2)}{3\,R^2\,Q^2}\,x
G^{\supDGLAP}(x,Q^2)\,\,>\,\,1\,\,,
\eeq
the predictions for all observables which we discuss in this paper
are  as follows:
\begin{enumerate}

\item $ F^{SC}_2(x,Q^2)\,\,\Longrightarrow\,\,\frac{N_c}{2
\pi^2}\,\sum^{N_f}_{1}\,Z^2_f\,Q^2\,R^2 \,\ln \kappa_Q(x,Q^2)\,\,;
$

\item $ \frac{\partial F^{SC}_2(x,Q^2)}{\partial
\ln(Q^2/Q^2_0)}\,\,\propto\,\,R^2  Q^2 $ for $x \,\leq\,x_0$, where 
$x_0$ is determined  from the  equation:

$\kappa^{\supDGLAP}_G (x_0, Q^2)\,=\,1\,\,.$   The   solution    to the
equation $\kappa^{\supDGLAP}_G (x, Q^2)\,=\,1$ is plotted in Fig.4, where we
plot also the solutions for $\kappa^{\supDGLAP}_G\,=$\,0.5 and 1.5.

It should be stressed that at any value of $Q^2$,  for very small $x
\,<\,x_0$ ,  the limit of the  $F_2$ slope  $ \frac{\partial
F^{SC}_2(x,Q^2)}{\partial
\ln(Q^2/Q^2_0 )}$
is   independent of $x$,  and is proportional to $Q^2$.

\begin{figure}[htbp]
\begin{center}
\epsfig{file=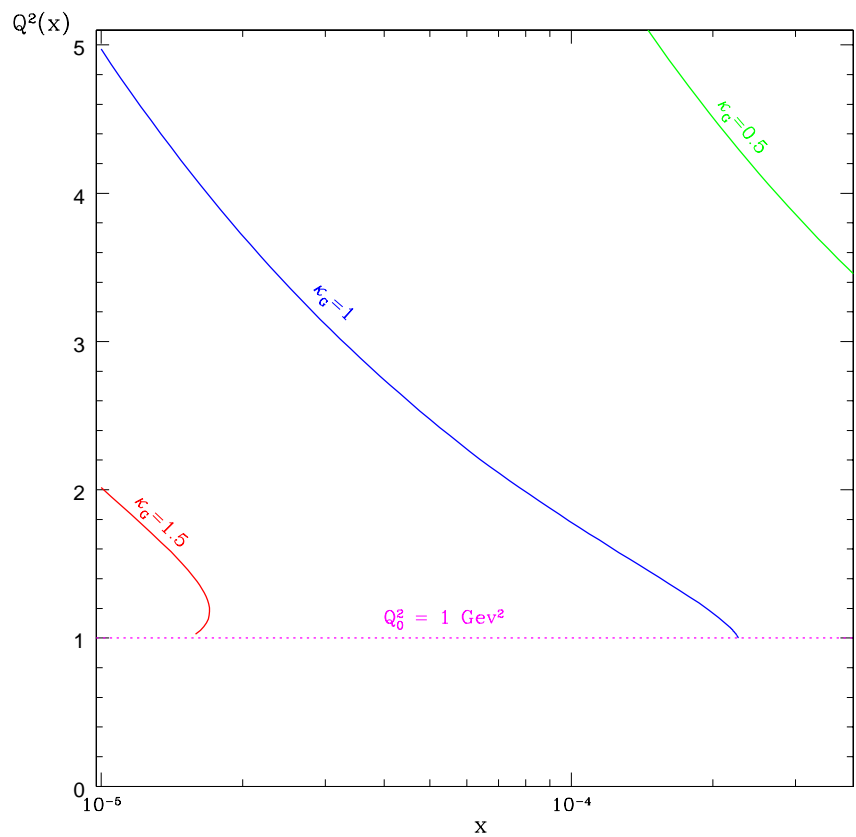,width=120mm}
\end{center}
\caption{Solutions to $\kappa_G\,=$ 0.5, 1 and 1.5.}
\label{Fig.4}
\end{figure}

 This limiting behaviour is  reached at smaller and smaller $x$ values as
$Q^2$ is increased, and it may explain   the $Q^2$ - dependence of the
experimental data ( see Fig.1 )\,\,;

\item  $ \frac{\partial \ln F^{SC}_2(x,Q^2)}{\partial
\ln(1/x)}\,\,\propto\,\,\frac{1}{\ln (1/x) \,\ln \ln (1/x)}$\,.
Therefore, in the region of ultra small $x$:\,\,   $\lambda (Q^2)\,=\, 
\frac{\partial \ln F^{SC}_2 (x,Q^2)}{\partial \ln (1/x)}\,\,
\Longrightarrow\,\,0$,    modulo logarithmic corrections  
independent  of the value of $Q^2$. We shall elaborate on this limiting
behaviour in the next section\,\,;

\item $xG^{SC}(x,Q^2)\,\,\Longrightarrow\,\,\frac{2\,R^2 \,
Q^2}{\pi^2}\, \ln(1/x)\,\ln \kappa_G^{\supDGLAP}(x,Q^2)$.
Comparing this behaviour of the gluon structure function with the
asymptotic behaviour of  $ \frac{\partial F^{SC}_2(x,Q^2)}{\partial
\ln(Q^2/Q^2_0)}$ one can deduce     that \eq{1} does not hold in the
region of
exceedingly small $x$.
 
\end{enumerate}

Obviously, the above are just the  exceedingly small $x$ limits  of the
actual
expressions, but
they
give us, together with the DGLAP predictions, a  framework to discuss
the experimental results,  so as  to develop a strategy of
measurement which can test  the asymptotic predictions. 

\section{Comparison   with the experimental results \\  and pseudo
data}
The calculations   presented in this paper  were carried out with
$R^2\,=\,10\,\gev2$ and $Q^2_0\,=\,0.4\,\gev2$. We have not attempted to
produce a ``best fit". However, we have satisfied ourselves that with
these input parameters, we obtain good  results which
provide a consistent  reproduction of the measured and pseudo  data. As
stated,
we have not pursued  further  the study of a two radii model
\cite{GLMSLOPEF} \cite{GLM403} being unable to suggest a decisive
signature  which supports this hypothesis.

\begin{figure}[htbp]
\epsfig{file=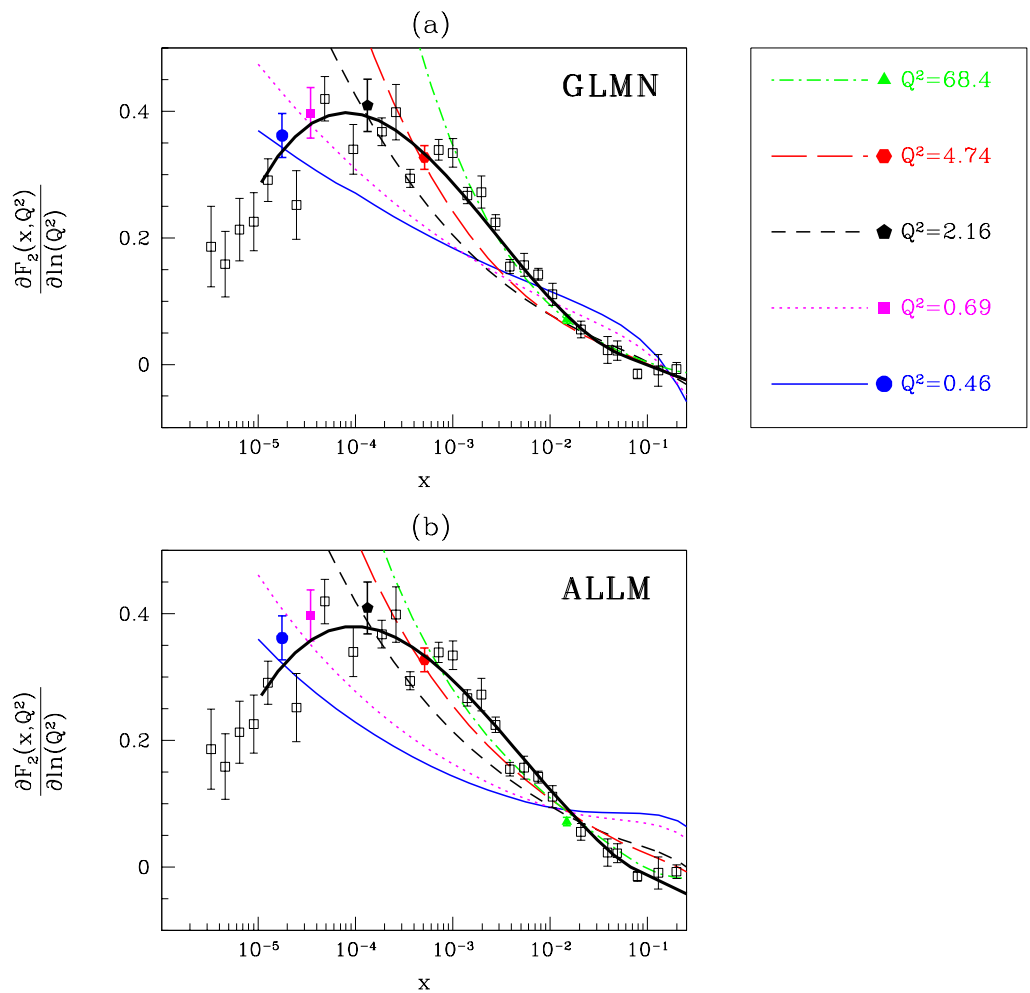,width=140mm}
\caption[GLMN and ALLM calculations for ZEUS data]{\parbox[t]{
    0.80\textwidth}{ (a) Our
    calculations for ZEUS data.  The thick line is the prediction for
    the actual ZEUS data which were taken at values of $Q^2$
    correlated with the value of $x$. (b) The same with ALLM'97 .}}
 \label{Fig.5}
\end{figure}

The detailed features of our reconstruction of the ZEUS data
\cite{DATA2} are shown in Fig.5a. As noted, the confusing compilation
of experimental data points with widely spread ($x,Q^2$) values
presented in this figure, makes it difficult to directly assess the
$x$ dependence of $ \frac{\partial F_2(x,Q^2)}{\partial \ln (Q^2/
Q^2_0)}$ at a fixed $Q^2$.  Our analysis suggests that $
\frac{\partial F_2(x,Q^2)}{\partial \ln (Q^2/ Q^2_0)}$ grows
monotonically with $\frac{1}{x}$ at fixed $Q^2$, approaching its
limiting value at $x$ which is well below the $x$ interval valid in
our calculations. At high $Q^2$ and at fixed $x$, this rise is steep
following the behaviour of $xG^{\supDGLAP}( x,Q^2)$ as expected from
\eq{1}. In our approach, this is a direct consequence of the fact that
in the high $Q^2$ limit and at fixed $x$,
$D^2_Q\,\,\approx\,\,D^2_G\,\,\approx\,\,1$ throughout the $x$ range
of interest. Generally, for any fixed $Q^2$ we can find $x_0$
sufficiently small so that for $x\,<\,x_0$ SC are important and, thus,
both $D^2_Q $ and $D^2_G$ are significantly smaller than 1. However,
from a practical point of view, as data are available for
$x\,>\,2\times 10^{-6}$ and the GRV'94 parameterization is defined for
$x\,>\,10^{-5}$, SC become relevant only at moderate and small $Q^2$
values ( see Fig.3 ).  This general behaviour is reflected in our
calculations where $ \frac{\partial F^{SC}_2(x,Q^2)}{\partial \ln
(Q^2/ Q^2_0)}$ departs from the DGLAP predictions, and as $Q^2$
decreases the $x$ dependence of $ \frac{\partial
F^{SC}_2(x,Q^2)}{\partial \ln (Q^2/ Q^2_0)}$ becomes more moderate.
Note that our approximations are not valid for $x\,>\,10^{-1}$ values.
The final compilation of the calculated points, at $( x, Q^2 )$ values
matching the experimental ones, reproduces the data very well, as is
evident from the thick line of Fig.5a. The other lines in Fig.5a are
for five typical constant $Q^2$ values, selected from the experimental
data. Note that the thick curve in the figure, like the data points,
belong to different $Q^2$ values.

In Fig. 5b we show the same plot as in Fig.5a, only this time the
fixed $Q^2$ behaviour is determined from ALLM'97 \cite{ALLM97}.
Evidently, the behaviour of $ \frac{\partial F_2(x,Q^2)}{\partial \ln
(Q^2/Q^2_0)}$, at fixed $Q^2$, shown in Fig.5b is very close to our
calculated behaviour presented in Fig.5a.

 Assuming the ALLM'97 pseudo data to be a reliable reproduction of
the, unmeasured yet, real data - we can assess the various theoretical
ideas and predictions for $ \frac{\partial F_2(x,Q^2)}{\partial \ln
Q^2}$, which are compared with the pseudo data in Figs 6 and 7.

\begin{figure}[tbp]
\epsfig{file=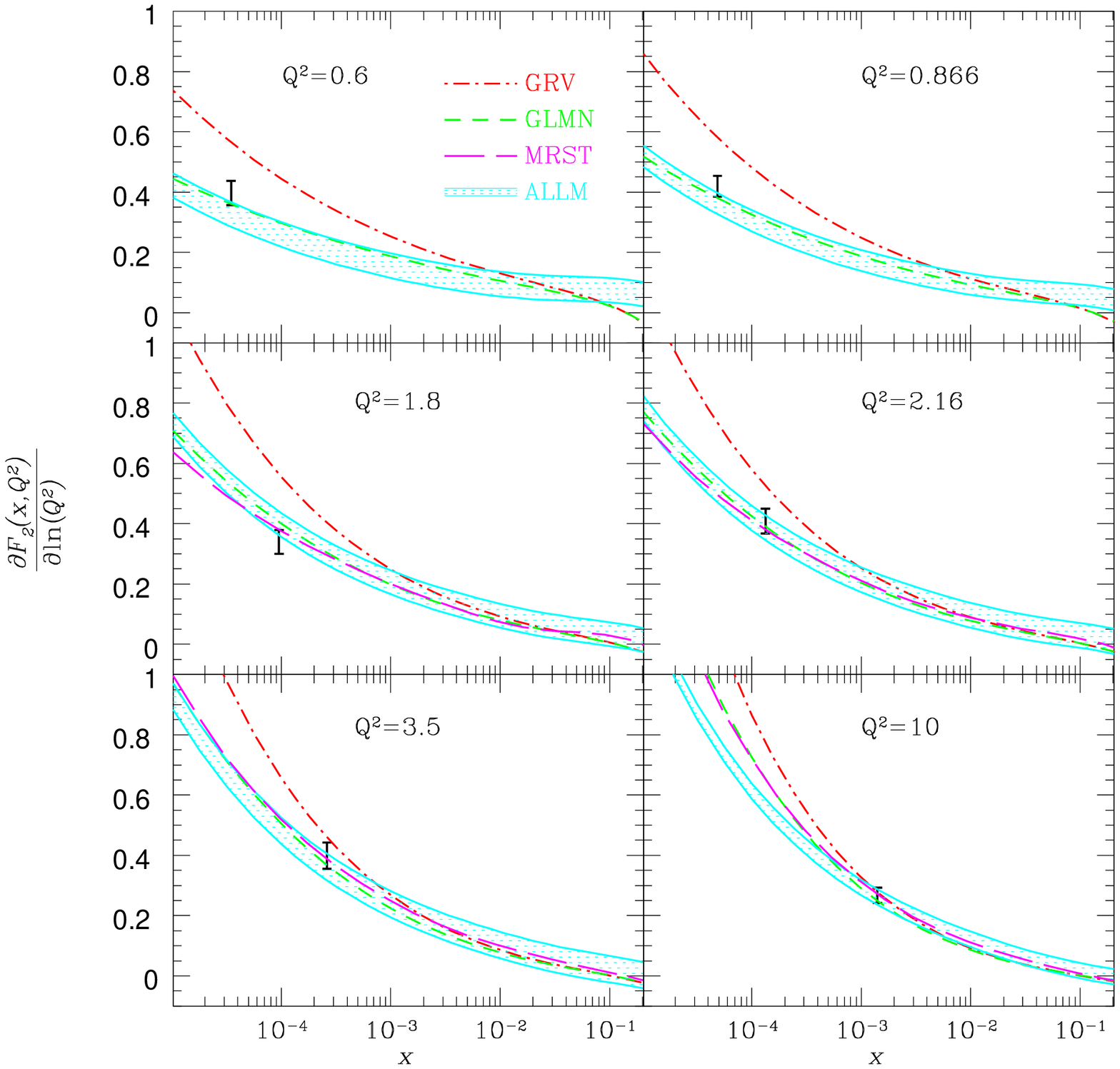,width=140mm}
\caption[$\df2dlnq2$ at fixed $Q^2$]
        {\parbox[t]{0.80\textwidth}{$\df2dlnq2$ at fixed $Q^2$. In
        addition to the ALLM band we show a typical data point with
        its error}}
\label{Fig.6}
\end{figure}

\begin{figure}[tbp]
\epsfig{file=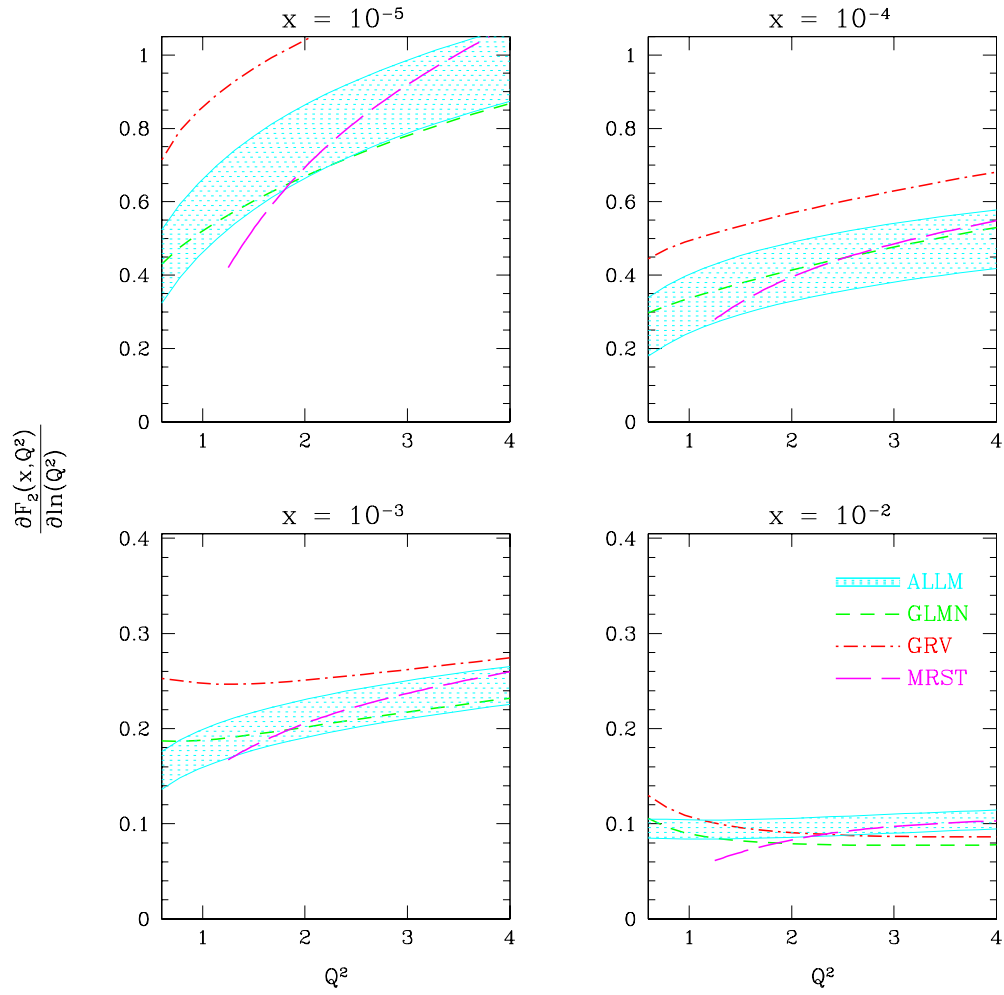,width=140mm}
\caption{$ \frac{\partial F_2(x,Q^2)}{\partial \ln Q^2}$ at fixed $x$.}
\label{Fig.7}
\end{figure}

A detailed comparison between ALLM'97 and the results obtained from
our calculation ( GLMN ) as well as GRV'94 \cite{GRV} and MRST
\cite{MRST} are presented in Fig.6 for various $Q^2$ values. ALLM'97
is presented as a band which includes the estimated error
\cite{ALLM}. We also show the typical small $x$ experimental errors,
so as to better assess the quality of the theoretical predictions.  In
Fig.7 we compare the pseudo data $Q^2$ dependence of $ \frac{\partial
F_2(x,Q^2)}{\partial \ln Q^2}$ at fixed $x$ values of $10^{-2}$,
$10^{-3}, \,10^{-4}$ and $10^{-5}$ with the same parameterizations,
and the results of our calculation.  Note that the MRST \cite{MRST}
parameterization is valid only for $Q^2\,>\,1.25\,\gev2$.
\begin{enumerate}
\item Figs.5 - 7 show that the SC are able to account for the
deviations from the DGLAP prediction, observed experimentally and
reproduced in the ALLM'97 pseudo data.  Our calculations are
compatible with the observable scale $Q^2\,\approx\,2 - 4 \,\gev2 $
where such deviations start to be visible.

\item  In our approach
we do not see any indication supporting an abrupt  transition 
from the predominantly ``soft" region to the hard one.
 This is compatible
with the ALLM'97 fit  which had  to
choose a high Pomeron mass scale $m^2_P\,\approx\,50\,\gev2$, so as to
obtain a very gradual increase of the effective $\alpha_P(0)$, with
increasing $Q^2$ as required by the data. 

\item Figs.5 - 7 are compatible with a   new scale of hardness ( $Q^2
(x) $ ) suggested  in our main formulae ( \eq{17} - \eq{21} ).
$Q^2(x)$ is the solution of  the equation 
\beq \label{23}
\kappa^{\supDGLAP}_G (x, Q^2(x))\,=\,1\,\,.  
 \eeq
The general features   of this equation are shown in Fig.4, where the
relevant $Q(x)$ values can be identified. The novelty of this approach is
that a hardness scale is derived for any $x$ value under consideration. 
\end{enumerate}

Comparing GLMN with GRV'94 and MRST we observed that both GLMN and MRST
are significantly different from GRV'94 in accord with the   real and
pseudo data. We note  also that in the exceedingly small $x \approx 
10^{-5}$
region the $Q^2$ dependence of MRST becomes  different from ours but
we are unable, at this stage, to check it experimentally.

A summary of the $Q^2$  dependence of 
$\lambda( Q^2 )\,\,=\,\,\frac{\partial \ln  F_2(x,Q^2)}{\partial \ln (1/x)}$  
at fixed $x$ values of $10^{-5}, \,10^{-4}$ and $5\times 10^{-3}$  is
presented in Fig.8, where, once again, the ALLM'97 pseudo data is
compared with the GRV'94   and MRST parameterizations and our
results. The GLMN calculation was carried out with
$\mu^2\,=\,0.6\,\gev2$ 
controlling our continuation to very small 
$Q^2\,<\,0.6\,\gev2$, where 
$xG(x,Q^2 < \mu^2)\,=\,\frac{Q^2}{\mu^2}xG(x,\mu^2)$. 
To check our sensitivity we plot in Fig.9 the results of several sets
of calculations where in GLMN0 we have 
$xG(x,Q^2 < Q^2_0) = 0$ and $Q^2_0\,=\,1\,\gev2$, in GLMN1 we repeat
the same calculation  putting the GRV initial condition for $x G(x,Q^2_0)$
at  $Q^2_0\,=\,1\,\gev2$. 
GLMN is as in Fig.8 for three different values of
$\mu^2$: $\mu^2 = 0.4,\, 0.6\,\, \mbox{and}\,\, 1.0\; \mbox{GeV}^2$.

The study of $\lambda(Q^2)$ can be perceived from different points of view:
\begin{enumerate}
\item  In soft Regge approach we have, in the small $x$ limit, $F_2
\,\propto\,x^{- \lambda}$ where
$\lambda\,\,=\,\,\alpha_P(0)\,-\,1\,\approx\,0.08$.
$\alpha_P(0)$ is the ``soft" Pomeron trajectory intercept  at $t = 0$. 
In Regge approach the Pomeron intercept, and $\lambda$, 
depend on $Q^2$ and $x$. Experimentally, the data available for $ 0
\,\leq\,Q^2\,\leq\,200\,\gev2$ indicates a slow increase of $\lambda$ with
$Q^2$. This is also reproduced by ALLM'97 parameterization, where  the
increase   persists up to the highest   $Q^2\,\approx\,10^{4}\,\gev2$
described by this parameterization.  This behaviour of $\lambda( Q^2 )$
may be attributed to DIS dynamics or the existence of an additional
``hard" Pomeron or both.

\item The DGLAP evolution equations predict in the region of small $x$
that
\beq \label{24}
\lambda\,\, = \,\,\,\sqrt{\frac{N_c \alpha_S \ln(Q^2/Q^2_0)}{\ln(1/x)}}
\eeq
i.e. we expect 
  $\lambda$   to slowly increase with $ \ln Q^2$ and to slowly decrease
with $\ln(1/x)$.
\item In the limit of exceedingly small $x$,
 SC predict that  $ \lambda
\,\propto\,\frac{1}{\ln(1/x)}$ independent of $Q^2$.
As can be seen in Fig.9, this limiting behaviour of $\lambda$  is
reproduced in all the GLMN calculations where \eq{xglin} 
is used  for very small $Q^2 < \mu^2$. 
\end{enumerate}

Fig.8 suggests that for exceedingly small $x$, GRV'94 parameterization is
consistent with ALLM'97 version, whereas GLMN and MRST  are somewhat below
and above ALLM'97 predictions, respectively. Experimentally, there is no
data to compare with. Moreover, it is difficult to assess if these
are real differences   or just numerical artifacts at the kinematic edge
of the various calculations.
Note that both GLMN0 and GLMN1 agree with ALLM'97 for all $x$ values,
as shown in the figures.

In our calculation for Figs. 8 and 9 one has to 
exercise extreme caution when evolving in $1\,\leq\,Q^2\,\leq\,3 \,\gev2$
region, as the $x$ dependence of $\frac{\partial \ln F_2}{\partial 
\ln(1/x)} $ is not yet asymptotic.

\begin{figure}[htbp]
\epsfig{file=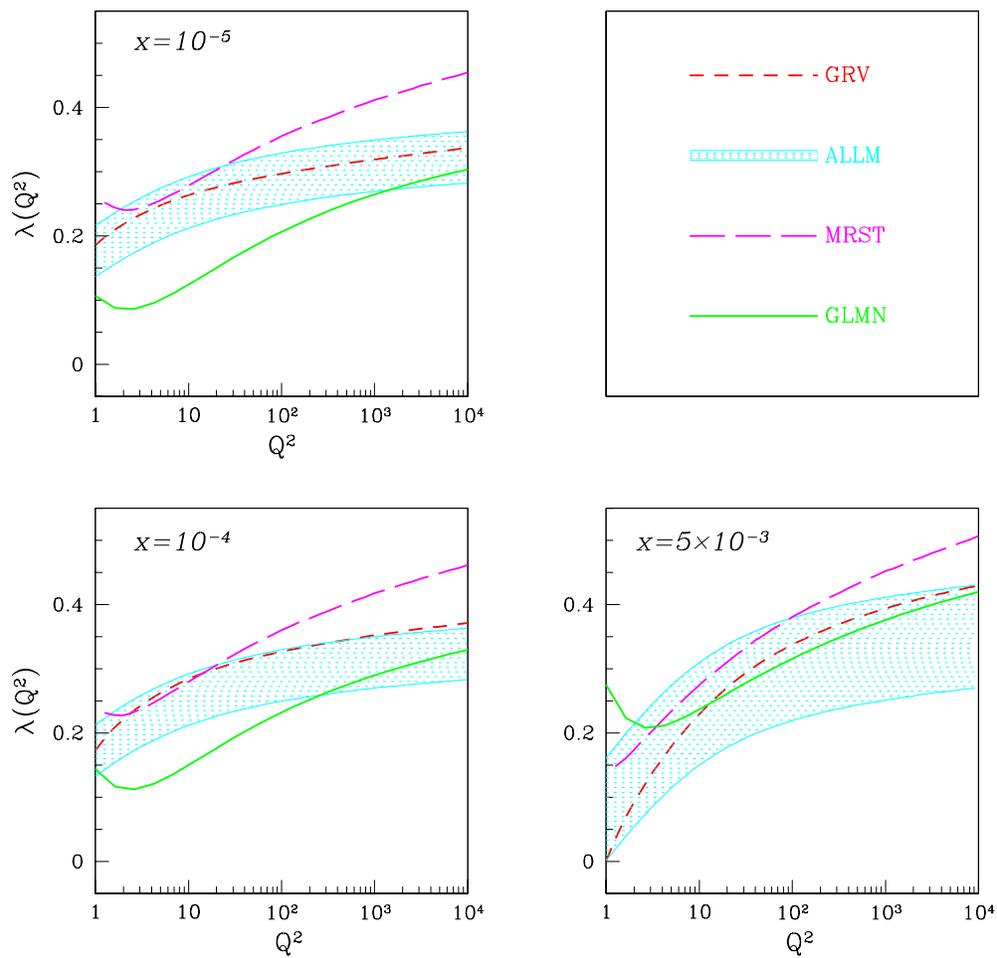,width=140mm}
\caption[Comparison of different approaches: $\lambda(Q^2)$ versus $Q^2$
at fixed $x$.]{\parbox[t]{0.80\textwidth}{Comparison of different 
    approaches: $\lambda(Q^2)$ versus $Q^2$ at fixed $x$ .}}
\label{Fig.8}
\end{figure}

\begin{figure}[htbp]
\epsfig{file=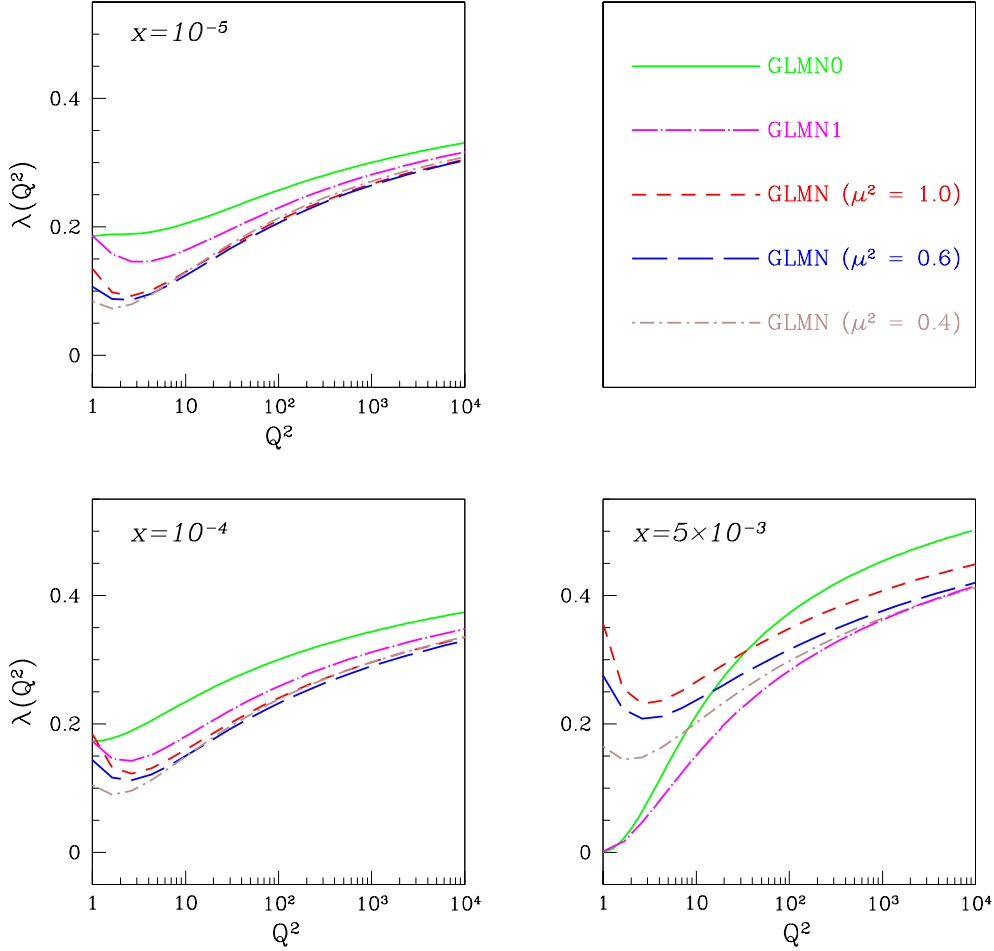,width=140mm}   
\caption[Comparison of different model for long distance behaviour of
$xG(x,Q^2)$ : $\lambda(Q^2)$ versus
    $Q^2$]{\parbox[t]{0.80\textwidth}{Comparison of different models
    for the long distance behaviour of $xG(x,Q^2)$: $\lambda(Q^2)$ versus
    $Q^2$ at fixed $x$ .}}
\label{Fig.9}
\end{figure}

\newpage 
\section{Summary and Conclusions}
In this paper we have attempted a detailed and systematic study of the
proton structure function logarithmic derivatives $\df2dlnq2$ and
$\dlnf2dln1x$. Our calculations are carried out in the double log
approximation of pQCD, including SC of the calculated quantities. The
SC are calculated in the Eikonal model. Our study was motivated by the
recent HERA results on $\df2dlnq2$ showing a considerable departure
from the DGLAP predictions in the kinematical domain of small $Q^2 < 5
\gev2$ and $x < 10^{-3}$.

A unique feature of our approach is that we compare our results with a
pseudo data base, which is computer generated from the ALLM'97
parameterization. This is done so as to overcome the lack of detailed
experimental data at sufficiently small $x$ and $Q^2$.  This enables
us to examine the fine details of our theoretical predictions, a task
which can not be accomplished with the limited relevant experimental
data presently available.

The main conclusions of our study are:
\begin{enumerate}
\item Our approach, in which we correct the unscreened DGLAP
  predictions for the effects of screening, enables us to achieve
  results on both logarithmic slopes of $F_2$, which are in good
  agreement with, both the real and pseudo data, at $x < 10^{-1}$.

\item The main feature of our investigation of $\df2dlnq2$ is shown in
  Fig.\ref{Fig.5} . It may appear that the particular structure of
  $\df2dlnq2$, evident in the ZEUS data, indicates that once $x$ is
  small enough then $\df2dlnq2$ changes from an increasing to a
  decreasing function of $\frac{1}{x}$.  This is {\em not} the structure
  suggested by our calculations. We demonstrated that $\df2dlnq2$ at
  fixed $Q^2$ both in our calculations, and in the pseudo data,
  remains a monotonic increasing function of $\frac{1}{x}$, in a good
  agreement with asymptotic expectations. The SC only suppress the
  {\em rate} of such growth in comparison with the DGLAP approach.  It
  is the combination of different $\left(x,Q^2\right)$ data points
  which creates the particular structure seen in ZEUS data, and
  reproduced in our calculation.

\item At low $Q^2$ we expect $\df2dlnq2$ to be proportional to $Q^2$,
  and therefore to decrease as $Q^2\rightarrow 0$. This result is in
  full agreement with our calculations, as is evident from
  Fig.\ref{Fig.7} . Comparison with the MRST parameterization shows
  that this effect can be reproduced in the evolution equation, by
  suppressing the value of gluon (quark) structure function at low $x$
  in initial partons distributions.

\item Since the damping of $\df2dlnq2$ is significant only for small
  $Q^2 < 5 \gev2$ and $x < 10^{-3}$, our calculations are on the
  boundary of being able to use pQCD. Moreover, this is a kinematic
  domain where both ``soft'' and ``hard'' dynamics are
  significant. Our calculations suggest a smooth transition between
  the ``soft'' and the ``hard'' domains, and as a consequence of this,
  the results that we obtain from ``hard'' pQCD calculations are shown
  to be stable, and compatible with the real and pseudo data. This
  observation is of a particular significance for the calculations of
  SC in the gluon sector, where our calculation also receives
  contributions from relatively long distances, for which pQCD
  provides no estimation of errors. We suggest that a scale of
  hardness be defined from the solution of \eq{22}. The above gives
  hope that the transition from "hard" to "soft" mechanism can be
  calculated theoretically. We consider this paper, together with our
  earlier Ref.\cite{GLMSLOPEF}, as a first attempt to quantify this
  description.

\item Our study of $\lambda\left( Q^2\right) = \dlnf2dln1x$ is
  compatible with both the real and pseudo data. We confirm the
  expected asymptotic behaviour of $\lambda$ in the region of very
  small $x$ (see Fig.\ref{Fig.9}). However, we would like to point out,
  that from Fig.\ref{Fig.8} one can see that the pseudo data indicate
  a different $x$ dependence of $\lambda(Q^2)$ than our
  calculations. Only real data at low $x$ can clarify the situation.

\item Our general approach was to start from the unscreened DGLAP, for
  which we assess GRV'94 to be most suitable, and then correct for the
  effects of screening. An alternative approach would be evolving
  parton distributions, as done in \cite{MRST}.  There are however, two
  main differences between our approach and MRST: the first is a
  practical deficiency of MRST, which is applicable only for $Q^2 >
  1.25 \gev2$ -- this limit is somewhat high for the present
  analysis. The second is the different predictions between our
  approach and the MRST parameterization, in the region of small $x
  \sim 10^{-5}$ (see Figs.\ref{Fig.6} -- \ref{Fig.9}).
\end{enumerate}

We firmly believe that the experimental systematic studies of the
$F_2$ slopes as well as other observables that are sensitive to the
value of SC ( $F_2(charm)$, $F_L$ , $\frac{\partial
F^{DD}_2(x,Q^2)}{\partial \ln Q^2}$\, etc.~), are needed to check one
of the most fundamental problem of QCD: the value of the scale of the
transition between perturbative and non-perturbative QCD.

{\bf Acknowledgments:} We thank H. Abramowitz and A. Levy for
providing us with a detailed program and for discussions on the
ALLM'97 parameterization.  This research was supported it part by the
Israel Science Foundation, founded by the Israel Academy of Science
and Humanities.


\begin{thebibliography}{99}
\bibitem{DATA1}
ZEUS Collaboration; J. Breitweg  et al.: \plb{407}{97}{432}; M. Derrick et
al.: \zpc{69}{96}{607},\zpc{72}{96}{394};\\
H1 Collaboration; S.Aid et al.: \npb{470}{96}{3},\npb{497}{97}{3}.
\bibitem{DATA2}
A. Caldwell: Invited talk in the DESY Theory Workshop, DESY, Octorber
1997.
\bibitem{DGLAP}
 V.N. Gribov and L.N. Lipatov: {\it   
Sov. J. Nucl. Phys.}{\bf 15}(1972)438;
 L.N. Lipatov: {\it Yad. Fiz.}{\bf 20} (1974) 181;
 G. Altarelli and G. Parisi: {\it Nucl. Phys.} {\bf B126} (1977) 298;
Yu.L.Dokshitzer: {\it Sov.Phys.  JETP} {\bf 46} (1977) 641.
 \bibitem{GRV}
 M. Gluck, E. Reya and A.
Vogt:  {\it Z. Phys.} {\bf C67} (1995) 433.
\bibitem{MRST}
 A.D. Martin, R.G. Roberts, W.J. Stirling and R.S. Thorne:
{\it ``Parton distributions: a new global analysis"}, DTP/98/10; {\tt
hep-ph/9803445}.
\bibitem{ALLM} 
H. Abramowicz, E. Levin, A. Levy and U. Maor: \plb{269}{91}{465}.
\bibitem{MRS}
A.D. Martin, R.G. Roberts and  W.J. Stirling: \plb{387}{96}{419}.
\bibitem{CTEQ}
CTEQ-collaboration: H.-L. Lai et al.:
\prd{55}{97}{1280};  J.Huston et.al.: {\tt hep-ph/9801444}.
\bibitem{GLR} 
L. V. Gribov, E. M. Levin and M. G. Ryskin: {\it Phys.Rep.} {\bf 100}
(1983) 1.
\bibitem{GLMSLOPEF}
E. Gotsman,E. Levin and U. Maor: \plb{425}{98}{369}.
\bibitem{BCF}
J. Bartels, K. Charchula and F. Feltesse: {\it Proceedings of the Workshop
`` Physics at HERA"} Oct.29 -30,1991, ed. W. Buchmueller and G. Ingelman,
v.1, p.193.
 \bibitem{ALLM97} 
H. Abramowicz and  A. Levy:{ \it ``The ALLM parameterization of
$\sigma_{tot}(\gamma^* p)$ an upgrade"}, DESY 97-251, {\tt
hep-ph/9712415}.
 \bibitem{DATA3}
   D,O, Caldwell et al.:                       \prl{40}{78}{1222}; 
   S.I. Alekhin et al.:                        CERN--HERA 87--01 (1987); 
   BCDMS Collaboration, A.C. Benvenuti et al.: \plb{223}{89}{485};
   L.W. Whitlow:                               SLAC--357 (1990);
   ZEUS Collaboration, M. Derrick et al.:      \zpc{63}{94}{391};
   H1 Collaboration, S. Aid et al.:            \zpc{69}{95}{27};
   E665 Collaboration, M.R. Adams et al.:      \prd{54}{96}{3006};
   NMC Collaboration, M. Arneodo et al.:       \npb{483}{97}{3}.
 \bibitem{LERY87}
 E.M. Levin and M.G.Ryskin:\sjnp{45}{87}{150}.
\bibitem{MU90}
 A.H. Mueller:\npb{335}{90}{115}.
\bibitem{REF11}
B.Z. Kopeliovich et.al.: {\it Phys. Lett.} {\bf B324} (1994) 469;
S.J. Brodsky et.al.:\prd{50}{94}{3134};
L.Frankfurt, G.A. Miller and M. Strikman: \prd{304}{93}{1};
 E. Gotsman, E.M. Levin and U. Maor:\plb{353}{95}{526};
 L. Frankfurt, W. Koepf and M.Strikman:\prd{54}{96}{3194};
A.L.Ayala Filho, M.B. Gay Ducati and E.M. Levin:  {\it Nucl. Phys.}
{\bf B493}(1997) 305; {\bf B510} (1998) 355;
 E. Gotsman, E. Levin and U. Maor: \npb{493}{97}{354}.
\bibitem{GLMPH}
 E. Gotsman, E.M. Levin and U. Maor: DESY 97-154, TAUP 2443-97, {\tt
   hep-ph/ 9708275}, { \it EPJ (in press)}. 
\bibitem{FT}
 J.C. Collins, D.E. Soper and G. Sterman: {\it Nucl.
Phys.}{\bf B308} (1988) 833.
\bibitem{AGLFR}
A.L.Ayala Filho, M.B. Gay Ducati and E.M. Levin: \plb{388}{96}{189}.
\bibitem{GLM403}
E. Gotsman, E. Levin and U. Maor: \plb{403}{97}{120}.
\bibitem{GLMD}
E. Gotsman, E. Levin and U. Maor: \plb{353}{95}{526}.
\bibitem{GLUONSC}
A.L.Ayala Filho, M.B. Gay Ducati and E.M. Levin:  {\it Nucl. Phys.}
{\bf B493}(1997) 305.
 \bibitem{GLMSMD}
E. Gotsman, E.M. Levin and U. Maor: \npb{493}{97}{354}.
\bibitem{MATH}
M. Abramowitz and I.A,Stegun: {\it ``Handbook of Mathematical
Functions"}, Dover, New York, 1970.
\bibitem{MUDIS98}
A.H. Mueller: { \it ``Small $x$  and diffraction scattering"}, plenary talk
at
DIS'98, Brussels, April 4-8,1998.  
\end{thebibliography}
\end{document}